\documentclass[12pt,preprint]{revtex4}
\usepackage{amsthm}
\usepackage{amssymb,dcolumn,amsmath}
\usepackage{bbm,mathrsfs,indentfirst,nicefrac,units}
\usepackage{graphicx}
\usepackage{multirow}
\begin{document}
\date{\today}
\title{ A dynamical network approach to uncovering hidden causality relationships in collective neuron firings}
\author{B\l a\.{z}ej Ruszczycki, Zhenyuan Zhao, and Neil F. Johnson}
\affiliation{Physics Department, University of Miami, Coral Gables,
FL 33146, USA}
\begin{abstract}
We analyze the synchronous firings of the salamander ganglion cells from the perspective of the complex network viewpoint where the network's links reflect the correlated behavior of firings. We study the time-aggregated properties of the resulting network focusing on its topological features. The behavior of pairwise correlations has been inspected in order to construct an appropriate measure that will serve as a weight of network connection.  
\end{abstract}
\maketitle

\section{Introduction}
The synchronicity between the firings of the optic nerve's ganglion cells might be a mean to encode the transmitted message. One of the reasons to believe that the coding actually takes place is the fact that the number of nerve fibers is much smaller than the number of photoreceptors in the vertebrate eyes, and therefore too small to transmit the signal directly without the loss of the resolution.

Here we use the dataset kindly supplied to 
us by Prof. Meister, in order to revisit the dataset of firings originally published in {\em Multineuronal Firing Patterns 
in the Signal from Eye to Brain}, Neuron {\bf 37}, 499Ð511 (2003) by Mark J. Schnitzer and Markus Meister. 
Our goal is to see if a complex network viewpoint may add new features to the analysis which appeared in this paper, and hence enhance our understanding of the collective dynamics of neurons.

Even if the coding scheme is dynamic (in the sense that it crucially depends on the way in which the observed picture changes in time)
its outcome is determined by the underlying structure whose certain properties should not change in time, the same coding scheme has to be maintained.
In principle, the most straightforward approach to understand the way in which the coding is formed would be to study the relation between the appearing multi-channel patterns of spikes as responses to appropriately selected visual stimuli. We are not at this point yet, the patch covered by the multi-electrode array is too small to gather enough information that allows to reconstruct the projected picture. There are however some static properties (characterizing the physiological system) that might be studied. One of these properties is the relation between the extent of correlation and the spatial structure, one may ask if the firing of some cell (under special external circumstances) is correlated with any other cell in the optical nerve or only with a limited set of them. This may reflect whether the coding is performed globally (in the extreme case the information from most of the cells is needed to understand the coded message) or locally, where there are only few cells whose actions may be correlated with the firing of a given one.  

One approach to analyzing the firing of cells is to consider the structure of groups of neurons -- or more specifically, nested sets since a single cell may participate simultaneously in many different groups. Here we take an alternative approach looking at the network whose links are determined by the coordinate firings of the cells. 
One may expect that both these approaches are related. In particular there are several techniques for hierarchical decomposition of a network, and the resulting hierarchy ultimately reflects the structure of the groups which would be created e.g. by minimizing the signal entropy. However, these two approaches highlight different features in different ways, hence the value of exploring the alternative dynamical network viewpoint. 

Coordinate behavior can give rise to intriguing features within a collection of objects, some of which we hope to capture in the present analysis. Consider an everyday scenario in which we explore the correlated appearance of three events A, B and C, by recording the
time interval between them each occurrence of certain pair. If A appears often together with B, B with C and C with A, it does not necessarily mean that A, B and C appear together at any one time -- hence in the dynamical network, all three pairs occur but they may not co-exist. If A, B and C do appear together, the three pairs appear simultaneously and hence a triangle now appears in the dynamical network. Now consider another scenario in which there is a "leader"  A and followers B, C, D... The appearance of A may be strong enough to attract the other members -- however B, C and D may not actually interact with each other directly. This would show up explicitly in the dynamical network as a hot-spot (i.e. preferential attachment). By contrast, this leadership role might be missed in the approach where the cells are assigned into a `group'. While the group viewpoint deduces structure based on a global measure, our network approach attempts to address global properties based on local measures related to causality.

In our particular case we do not mean by ``causality'' simply the triggering of one cell by the other (there are no electrical connections between the synapsis) but rather the following of one firing by the other as an effect of the common source. 

The proper treatment of the evolving network dynamics requires sufficiently long time-series in order to establish the results with acceptable statistical significance. Being restricted by this limitation we take a time-aggregated view of the causal dynamics, as explained below, in order to highlight any unexpected and persistent causal patterns. Specifically, our network will consist of directed links representing time-aggregated measures of the causal relationships. There are many possible quantities that might be used as the weight of the network links. This issue is discussed in Section~\ref{sec2}. 

\section{Pairwise measure}\label{sec2}
\begin{figure*}
\includegraphics[width=0.4\linewidth,viewport=200 0 1050 850,clip]{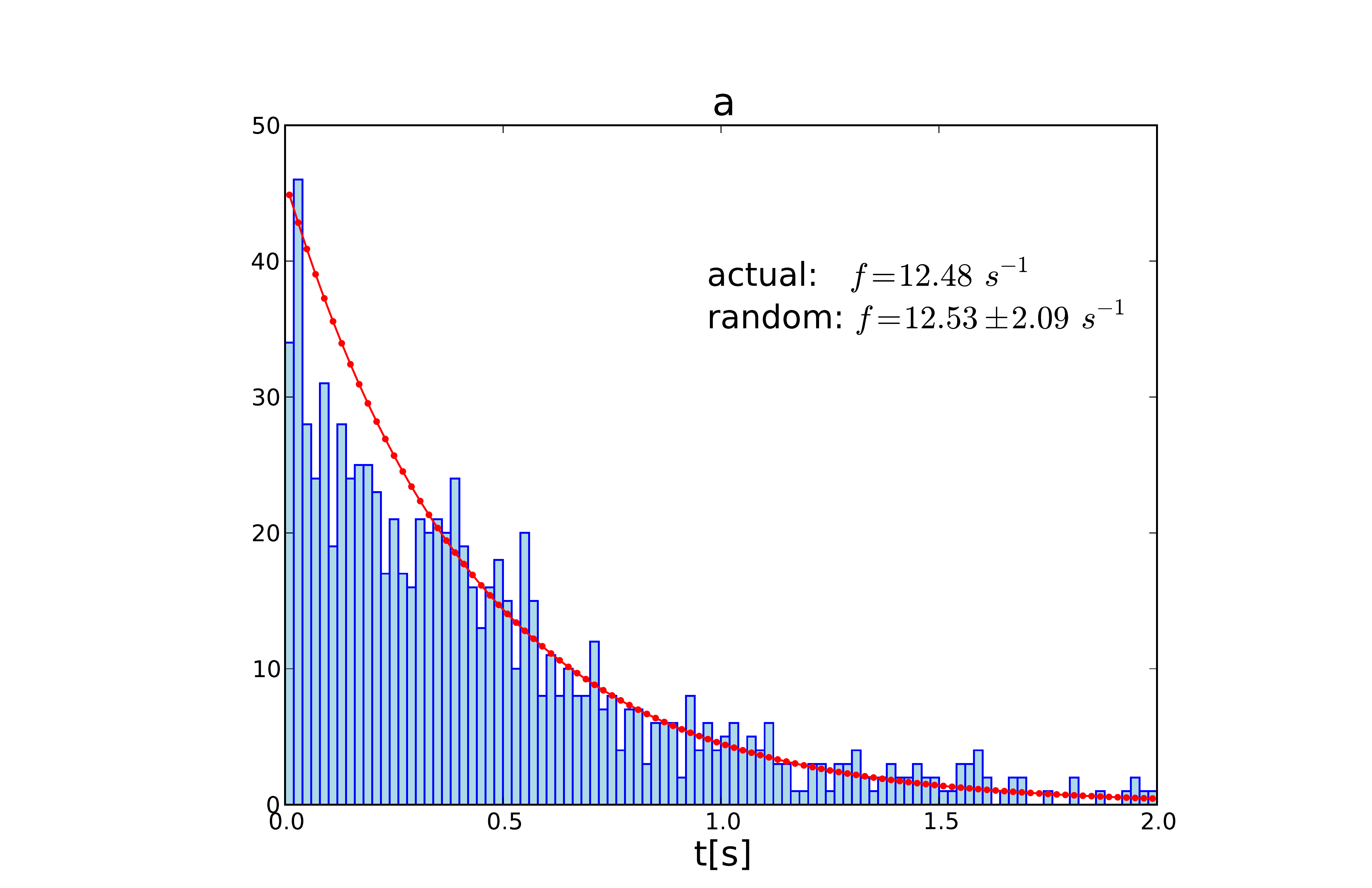}
\includegraphics[width=0.4\linewidth,viewport=200 0 1050 850,clip]{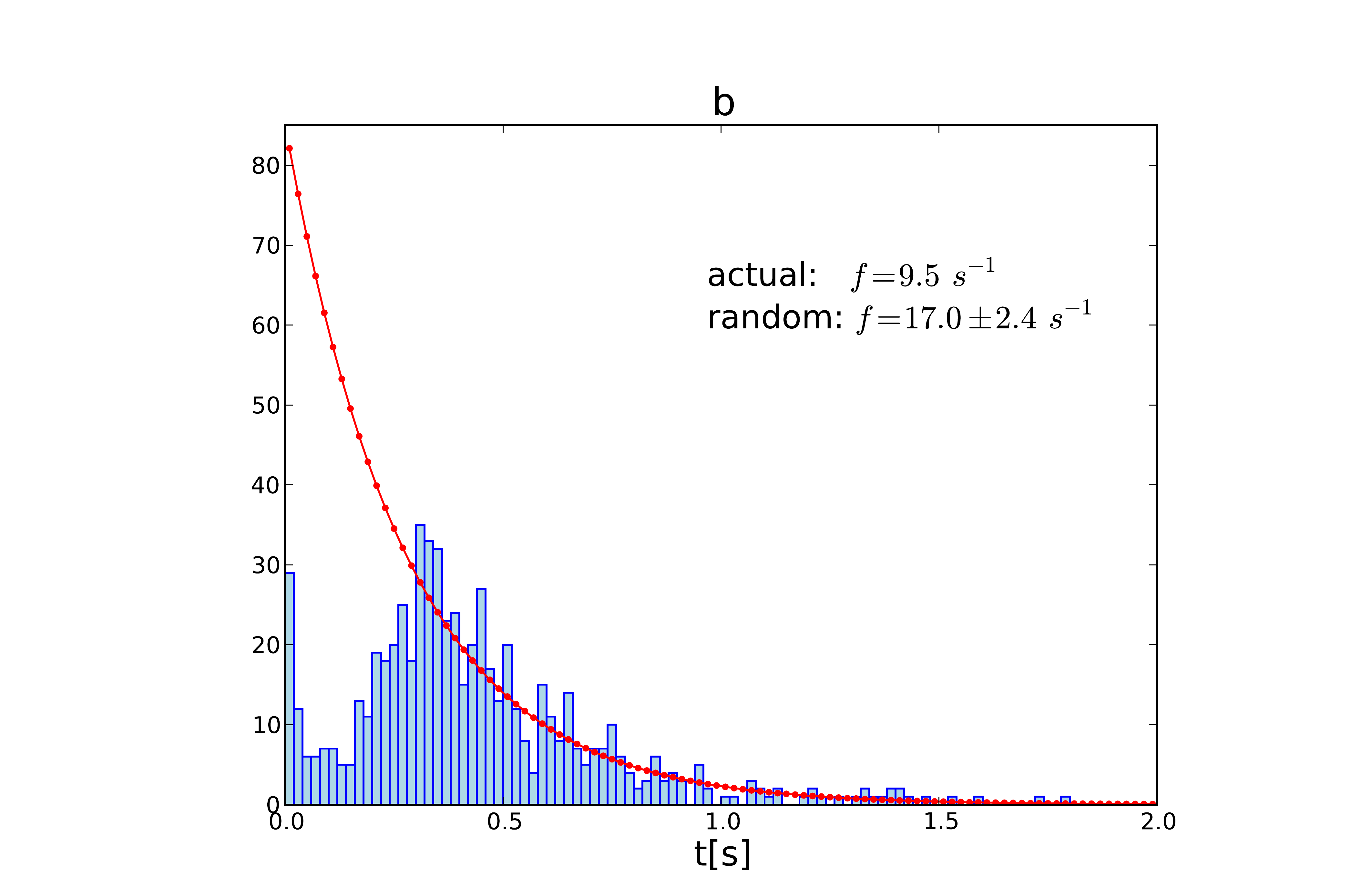}
\includegraphics[width=0.4\linewidth,viewport=200 0 1050 850,clip]{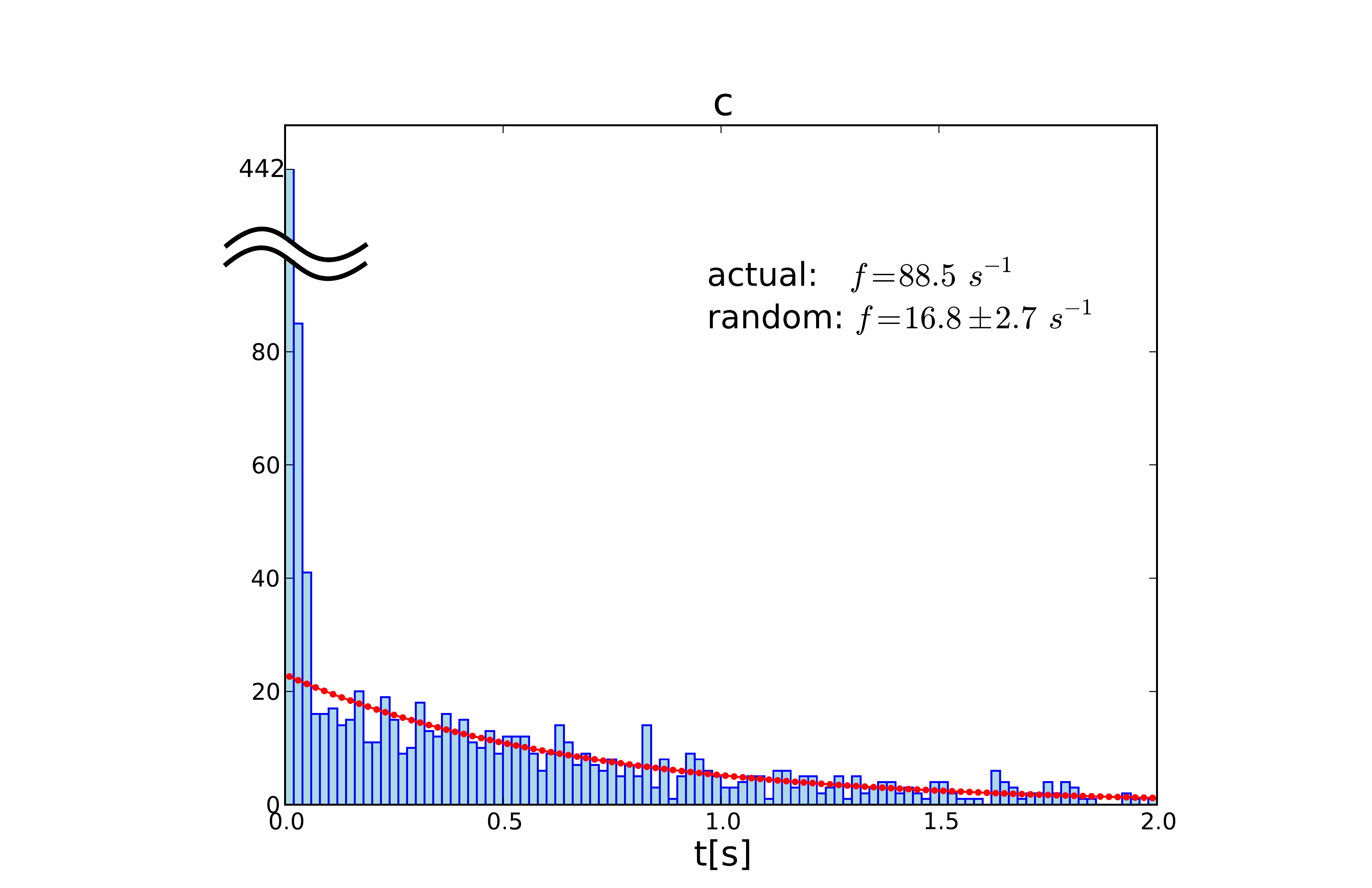}
\includegraphics[width=0.4\linewidth,viewport=200 0 1050 850,clip]{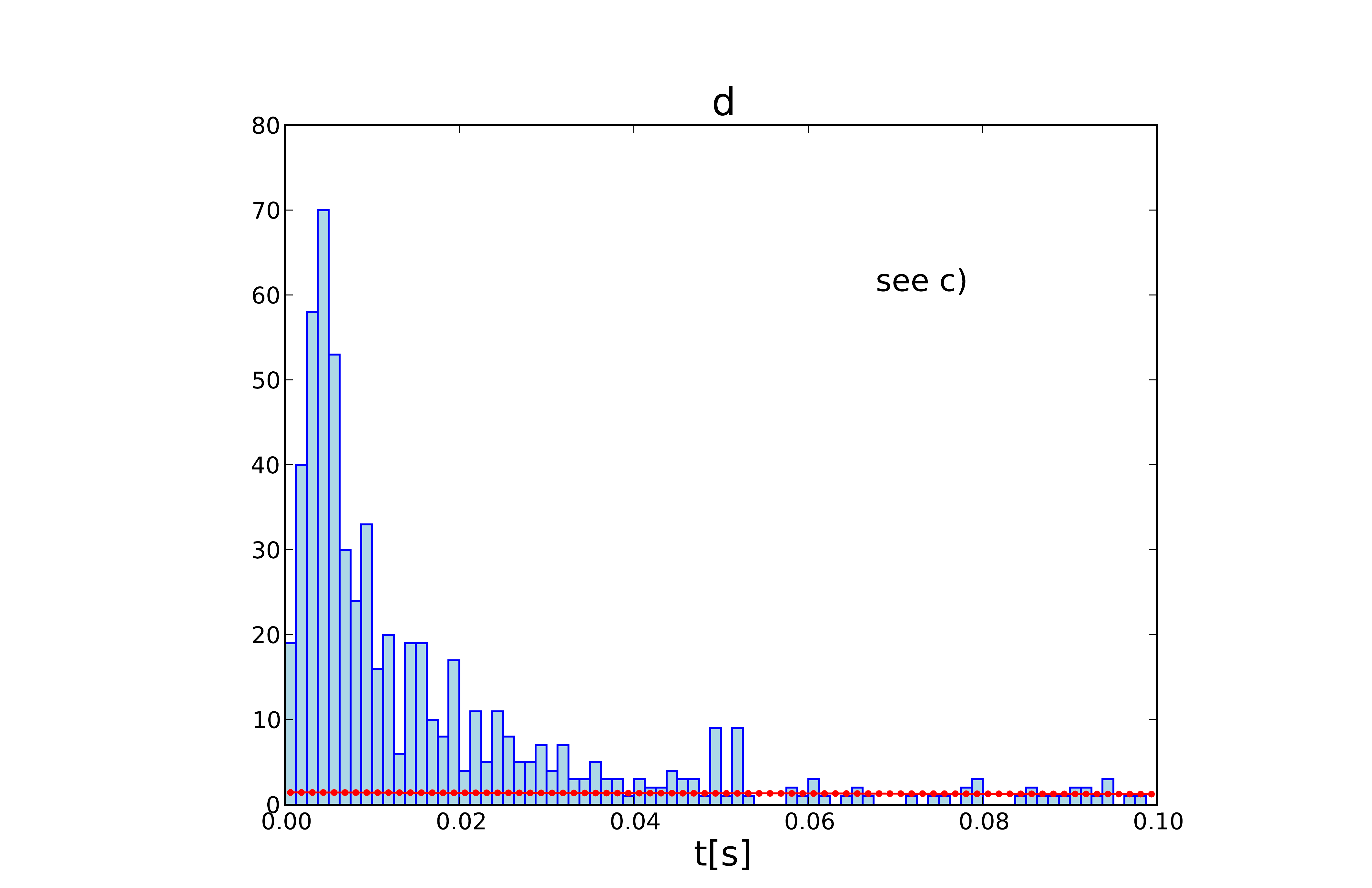}
\caption{Histograms of causal firing time for three example neuron pairs. We classify these as: a) random firing of B following A (i.e. no hidden causal relation); b) inhibition of firing of B following A; c) correlated firing of B following A. Histogram d) is a magnification of c) in the range $0-100\; ms$.
The red curve represents the Poissonian background evaluated by the estimator which was computed using the events for which $\Delta t>500ms$.} \label{f1}
\end{figure*}

 We abstract the ordering relationships between the firing of different cells, by looking at the time intervals between firings. For each pair of cells (e.g. A and B), we record the set of time intervals (i.e. waiting times) $\{t^i_{AB}\}$ between the firing of cell A and the next firing of cell B. We do this by combining the two time series of A and B and picking up the closest AB pairs. For example if cell A fires at time $t_A=1.1,2.5,..$ and the firings of cell B are at $t_B=1.2,2.9,..$, then these time intervals will be $t_{AB}=0.1,0.4,..$. We do not know if a given firing of A actually influenced the subsequent firing of B, but this is what we are aiming to find out. In order to decide what is the most appropriate pairwise dynamical measure between A and B, such that we can then produce a connection weight between A and B in the network, we stipulate the following requirements: 
\begin{itemize}
\item
Any non-recordings of spikes should influence the pairwise measure in a minimal way. As stressed in Ref. \cite{sg}, a large fraction of spikes are lost in the recording/recognition process which leads to a significant noisy background once we consider the distribution of time-intervals between the cross-channel  consecutive firings. See Fig.~\ref{f1}.
\item
The pairwise measure should take into account the significant variation of activity among different cells. The role of the pairwise measure is to highlight the hidden
causality between firings, not the more obvious relative frequencies of firing. We therefore have to be able to distinguish the case of two cells 
that rarely fire but do so in a coordinated way, from any accidental correlation of two very active cells.    
\end{itemize}

For the weight of a link between any two cells A and B, we propose to use the average of the reciprocal of the inter-firing time. The average value of $t_{AB}$ would be prone to the errors caused by non-recording of a spike, as discussed above. Moreover, we  
want to look at the causality among the firings by isolating the cases when the firing of a cell A is followed by a quick firing by a cell B. Specifically we will use the quantity $f_{AB}$ defined as: 
\begin{equation}
f_{AB}=\frac{1}{N} \left(\sum_{i=1}^N \frac{1}{t^i_{AB}+\epsilon}\right) \,.\label{eq_f}
\end{equation}
This quantity $f_{AB}$ differs from the average of $1/t^{i}_{AB}$ by the introduction of the constant $\epsilon$. Since we are interested in causality, and causality between two cells will ultimately rely on A's firing influencing B, it makes sense that there should be a minimum time interval for that influence to pass from A to B. This minimum time interval is the constant $\epsilon$. Without this, the divergent terms with $t \rightarrow 0$ would have given unphysically large contributions to the sum, with significant measurement errors. Thinking about typical firing windows, we choose $\epsilon=1\,ms$, but we have checked that for $\epsilon=5\,ms$ for example, the values of the link weights change by no more than few percent.

We wish to pick up causal events which would not have happened by chance, hence we say that a network link exists between A and B  {\em only} if its value differs more than three standard deviations from the average value constructed from a set of randomized samples. To follow this procedure, we need to produce a set of randomized samples {\em without} contaminating them with any individual cell-firing heterogeneities -- hence we keep the original distribution of inter-spike intervals on any single channel, but desynchronize different channels by randomly shifting and inverting the time series. Following this procedure, approximately $75\%$ of links are discarded, leaving a relatively sparse network containing arguably only meaningful links.  
Among the links which do survive this test, we find that the majority satisfy $f_{actual}>f_{random}$. However there are some cases for which $f_{actual}<f_{random}$. Different examples are presented in Fig.~\ref{f1}. For the case of `inhibition' for a given cell pair AB, there are fewer firings of B following firing of A that we would have expected from an estimate of the background. Most of these inhibition cases originate from a single cell.

\begin{figure}
 \includegraphics[width=0.9\linewidth]{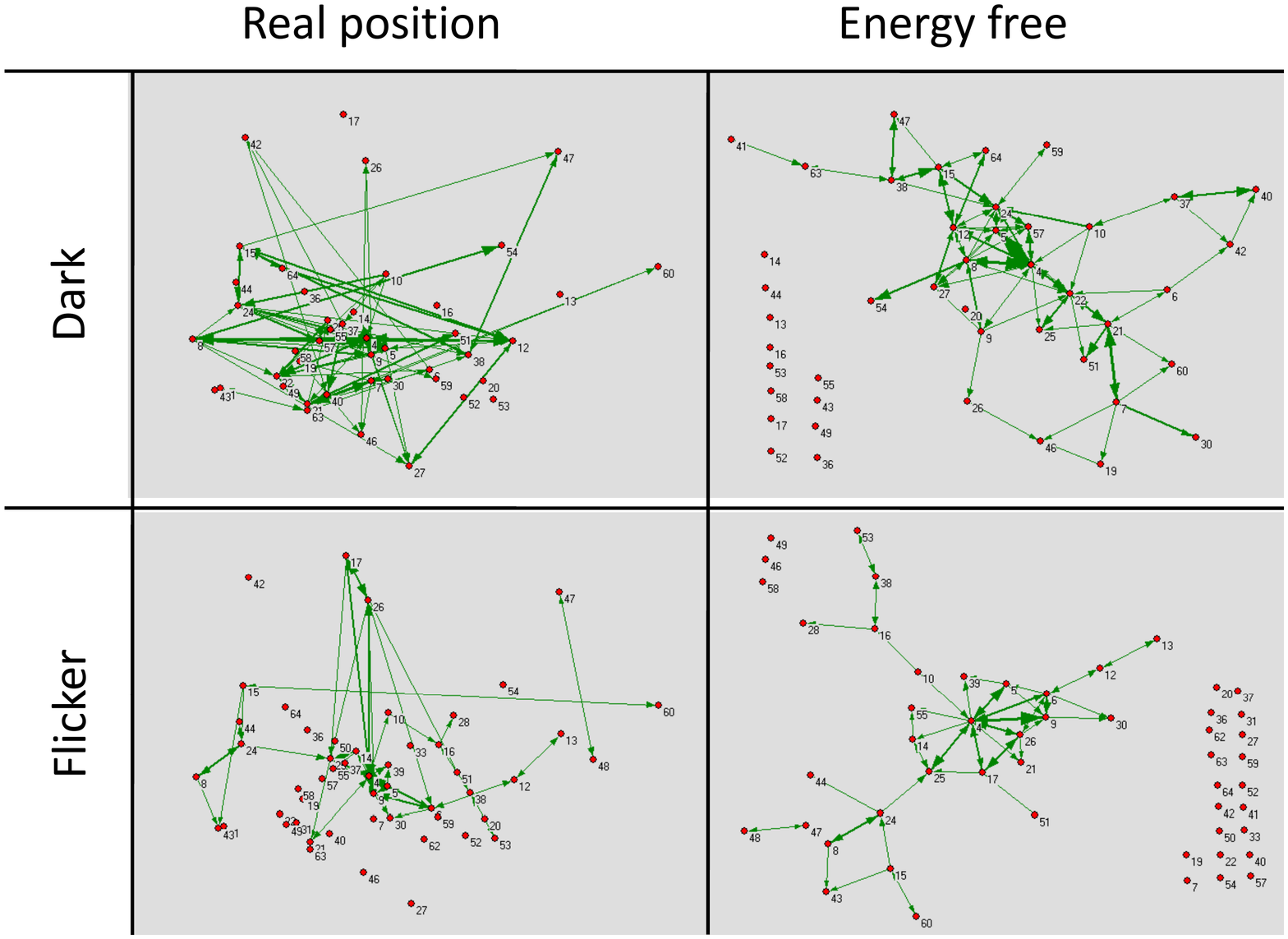}
\caption{The networks with spatial arrangements of the nodes. These `spatial' arrangements reflect the visual receptive field (see left) and the more abstract Kamada-Kawai algorithm (i.e. `energy free', see right) }\label{fig:network}
\end{figure}

\section{Network Properties}

Now that we have a set of links between the cells, we can construct a directed network with weights~$f$. For simplicity and illustrative purposes, we begin
the analysis by studying the network's strong component. We isolate this strong component by imposing the threshold of $f\geq60\,s^{-1}$ on the link weights. We note that the graph of the whole network containing $\approx 25\%$ would appear as a rather messy ``Gordian Knot''.

\begin{figure}
 \includegraphics[width=0.9\textwidth]{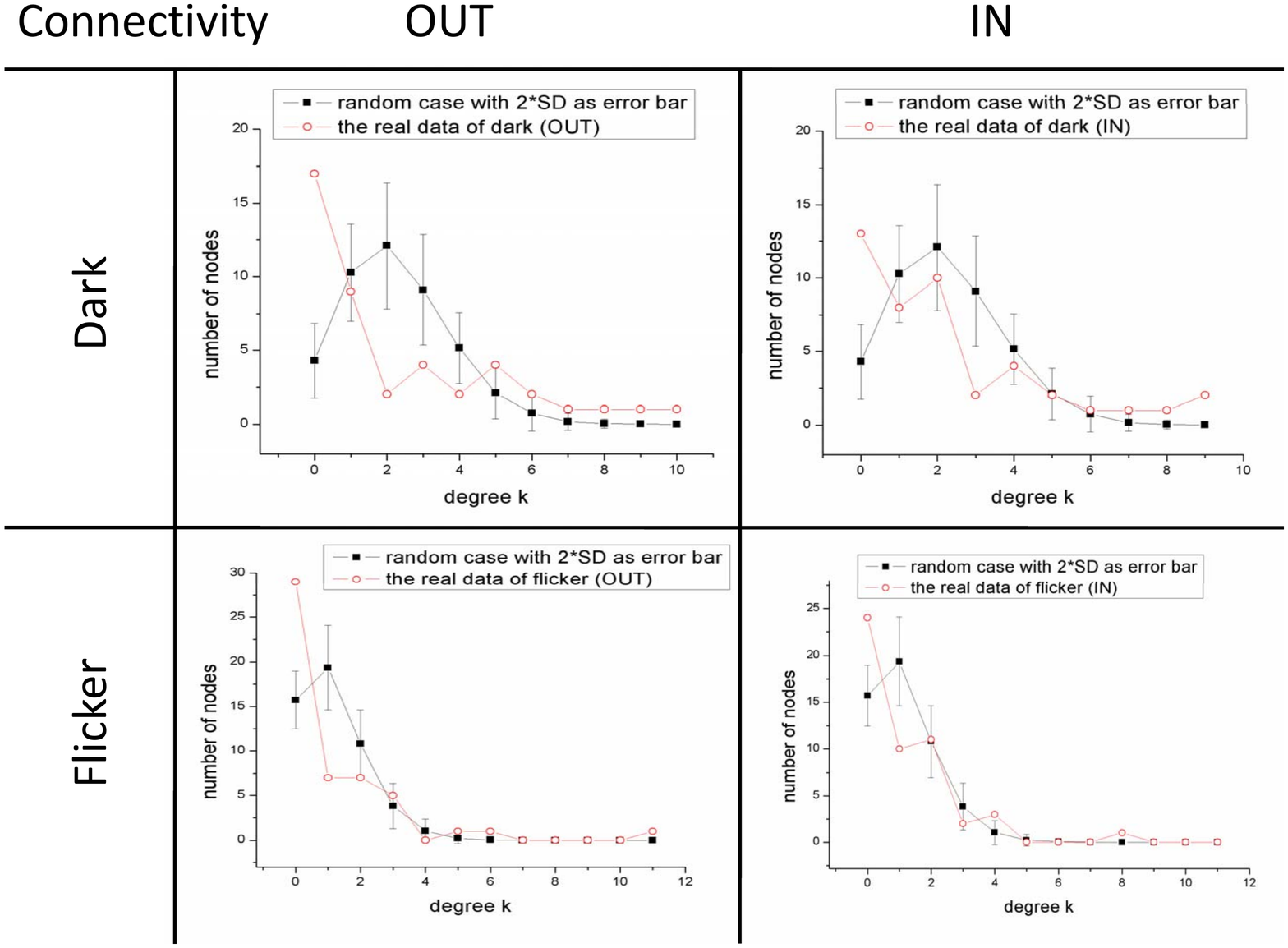}
\caption{The degree distribution of the actual data is compared to the one based on randomized samples (see text). The error bars represent here $\pm2$ standard deviations.}\label{fig:degree}
\end{figure}

The network's strong component is presented in Fig.~\ref{fig:network} where different spatial arrangements are presented. The spatial arrangement according to the visual receptive field is presented together with `energy-free' arrangement of nodes which was obtained by the Kamada-Kawai method, see~\cite{kk}.
By eye, we can see the suggestion of a clustering, or centering effect, in the middle of the graph around cell
No. 4 for both dark and flicker cases. The `real position' arrangements exhibit more crossings of the links than the `energy-free' arrangements, suggesting that the spatial position according to the visual field is {\em not} the important configuration dictating the correlations. We note however, according to Ref.\cite{ms}, that the values of these spatial coordinates may actually be biased by significant errors. We also note that the structure for the `Dark' case is more entangled than for the `Flicker' case.

\begin{figure}
 \includegraphics[width=0.4\linewidth]{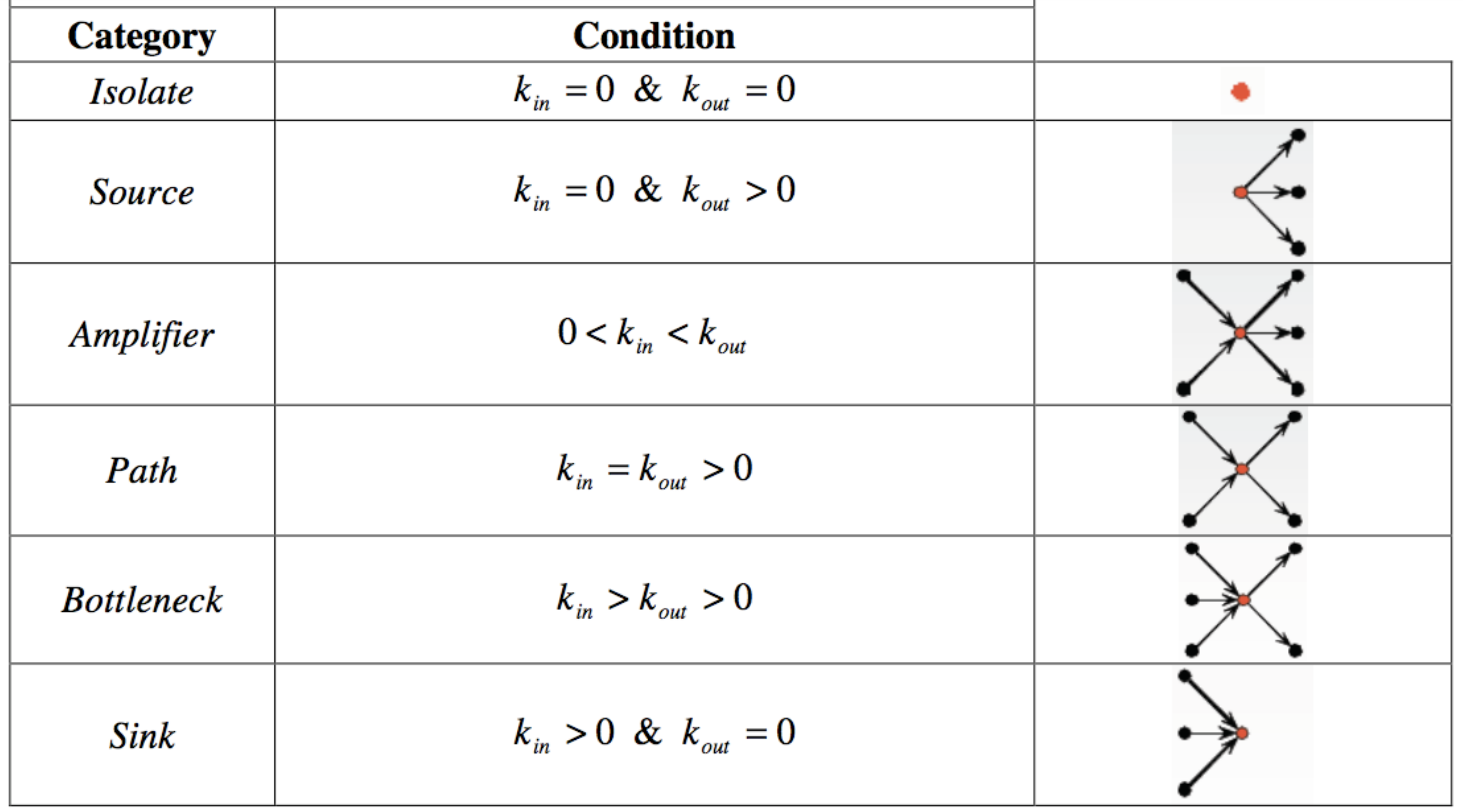}\\
 \includegraphics[width=0.2\linewidth]{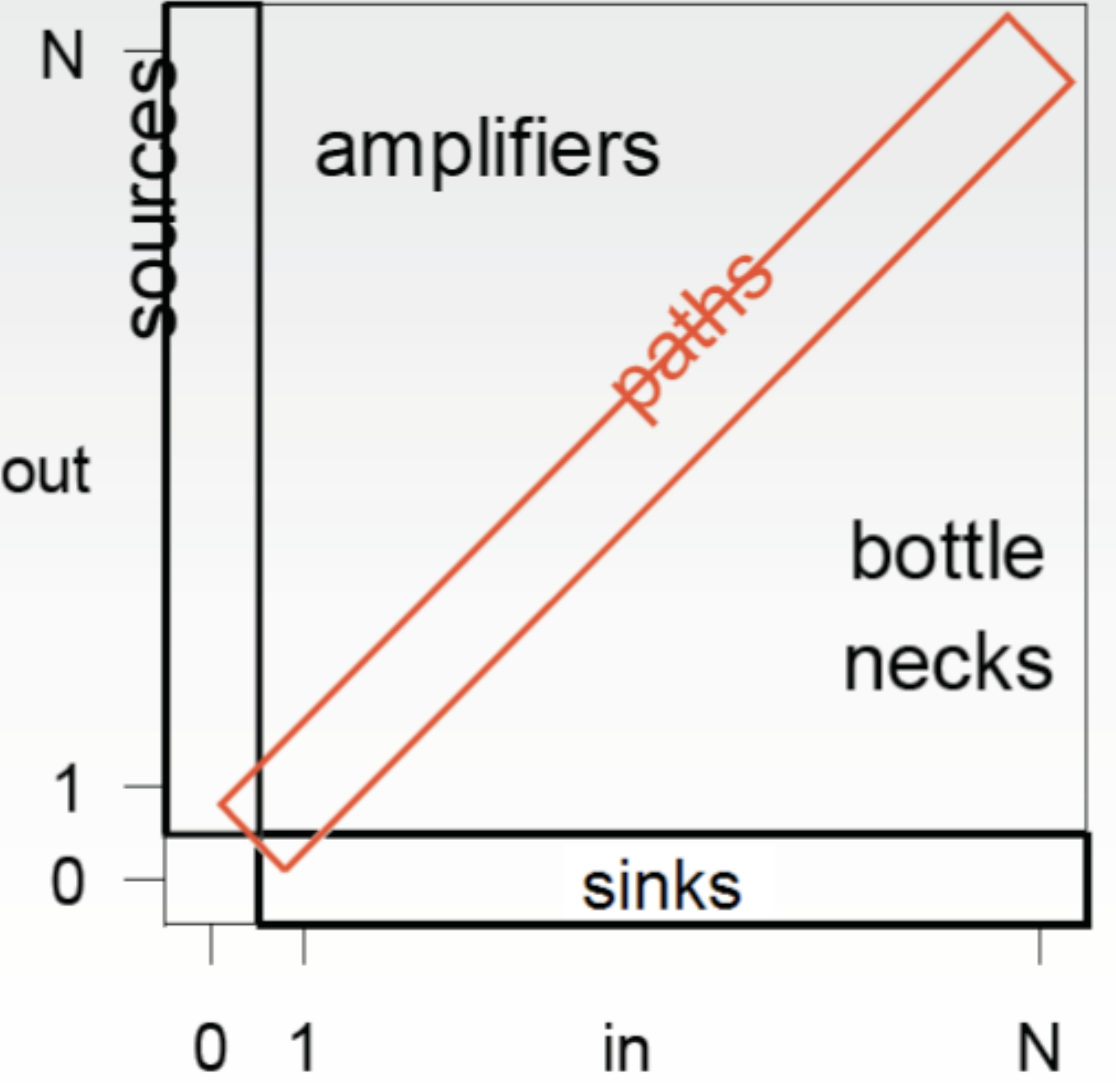}
\caption{Top: Schematic diagrams of motifs. Bottom: Maps of
motifs which are represented by different regions of the plot (see Fig~\ref{fig:zmap}).}\label{fig:motif}
\end{figure}

If we momentarily neglect the weights of links, considering them simply as existing or not, we can compare the connectivity
degree distribution $P(k)$ (the number of nodes with
$k$ links) with the random case, where the same number of links and
vertices is maintained but the links are randomly reassigned. This is a second form of randomization, allowing us to explore the structures beyond pairwise dependence since this randomization maintains the number of pairs, i.e. it maintains the number of links.
Fig.~\ref{fig:degree} shows that for firings recorded in darkness, the resulting network structure has an over-representation of zero connectivity (i.e. $k=0$) nodes, an under-representation of nodes with intermediate connectivity, and an over-representation of high connectivity nodes. For the firings recorded with `flicker light', there is no over-representation of high $k$ nodes present. This latter observation is consistent with the findings of Ref. \cite{ms} where there were more large groups detected for the `dark' case than for the `flicker' case.

\begin{figure}
 \includegraphics[width=0.75\linewidth]{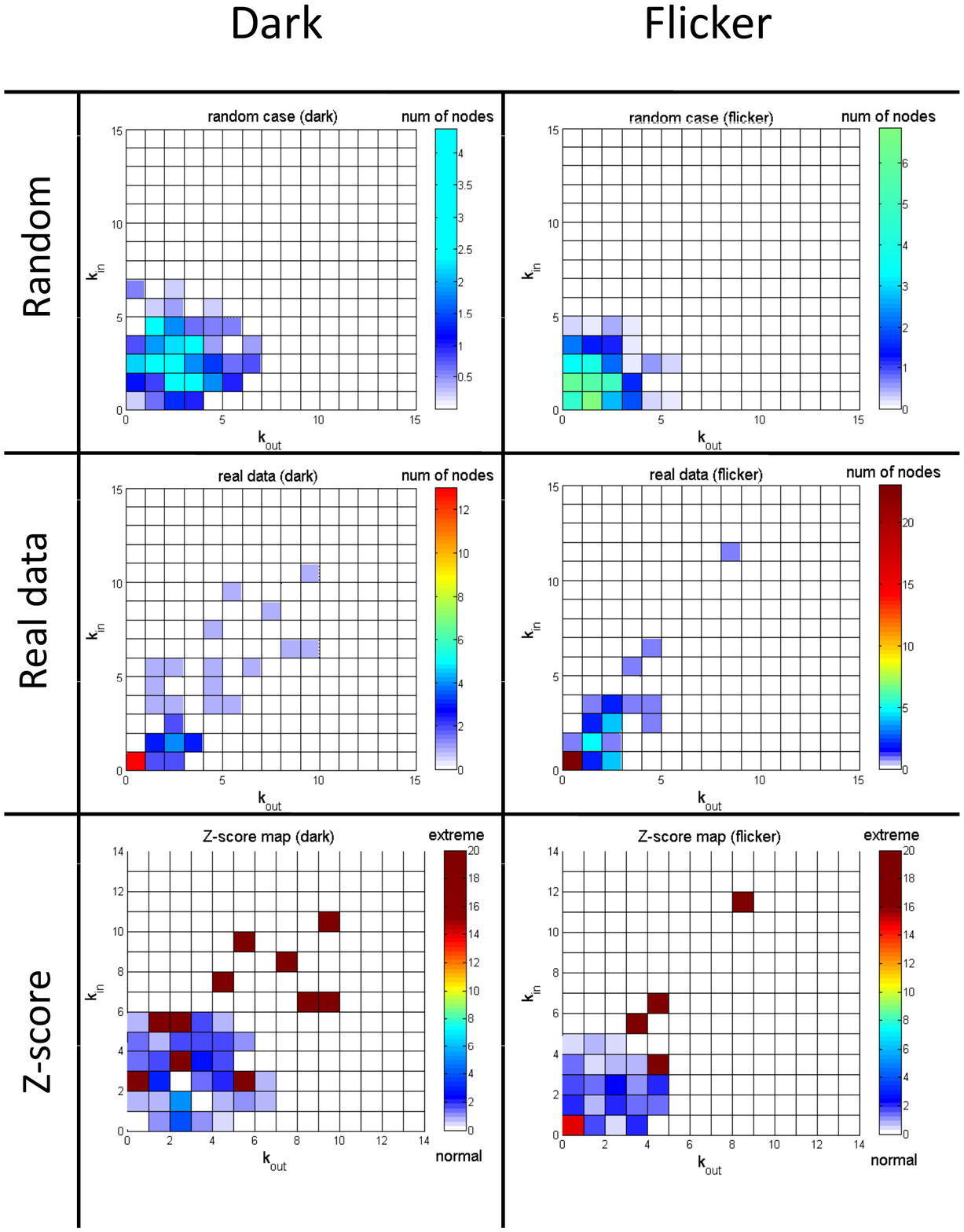}
\caption{Maps of motifs for the actual and randomized data. The last panel is the Z-score map, where the Z-score equals
$|x_{actual}-{\bar x_{random}}|{/\hat \sigma_{x_{random}}}$. The large values of the Z-score represent 
the abnormal configurations.}\label{fig:zmap}
\end{figure}

We now turn to look at $k_{in}$ (i.e. the number of incoming connections) and $k_{out}$
(i.e. the number of outgoing connections) for each node (i.e. cell) and hence will categorize
cells into different classes or motifs, as being either isolated, sources,
amplifiers, etc. See Fig.~\ref{fig:motif}, on the left, for a summary. Different cases
will map down to the different regions in
Fig.~\ref{fig:motif}, on the right.
Fig.~\ref{fig:zmap} shows a clear tendency for symmetry in the arrangement of links,
i.e. cells tend to have same amount of $k_{in}$ and $k_{out}$. In
other words, cells seem to concentrate on being `paths'. We also notice the deficiency of the `source' motifs, especially for the firings in the case of darkness.

\section{Discussion}

The striking peculiarity revealed by Fig.~\ref{fig:degree} and Fig.~\ref{fig:zmap} is that the structure of the time-aggregated network is much closer to the structure of the random network for the ``Flicker'' case rather than for the ``Dark'' case, as one may intuitively expect. This fact is not necessarily an argument against the transmission
of information by the mean of correlations. One possibility is that it is actually a lack of correlation that represents a specific message.
It might also have happened that reason for the randomness of ``Flicker'' case is data is the random checkerboard stimulus, if we think of the globally 
connected network (where there is little spatial-like arrangements) whose links are activated by incidence of light at specific points on the visual stimulus surface, we expect that the relations between the correlated pairs will be random once the activating source is such. This result is consistent with findings of \cite{ms} where the number of cells participating in groups is systematically smaller for the ``Flicker'' case than for the ``Dark'' one.

 Inspection of Fig.~\ref{fig:degree} does not show that there is a straightforward connection between the spatial structure defined by the visual stimulus and the network connectivity.
 
 We note also that that the coordinate action between the firings does necessarily have to reveal itself as a positive correlation, as seen in Fig.~\ref{f1} b. it may also have a form of ``suppression'' when the firing of one cell is blocked when the other one fires. Such behavior may have also a role in information coding and has to be included into the signal analysis algorithm. 

Bearing in mind the goal of uncovering novel dynamical collective behavior the subsequent path to explore would be
to consider the overall dynamical network structure. By this, we mean not just a spatial, static network, but a dynamical one in which the firing neurons form a highly fluid `soup' of groups whose number and membership may vary significantly over time.
The network analysis that we have done so far stops short of looking at the evolution of a dynamical network at each point in time. This could be looked at in the future in order to explore time-dependent causal patterns, however this would probably require much more data in order to produce reliable conclusions. We hope that the recent experimental advances will allow soon to have at the disposal such set of data.

\end{document}